\documentclass[11pt,twoside]{article}
\usepackage{asp2010}

\resetcounters

\begin{document}

\title{A constant characteristic mass for star forming galaxies since $z \sim 3$ 
revealed by radio emission in the COSMOS field
}
\author{Alexander~Karim, Eva~Schinnerer\\
\affil{Max-Planck-Institut f\"ur Astronomie,\\
K\"onigstuhl 17, D-69117 Heidelberg, Germany [email: karim@mpia.de]}
}
\author{and~the~VLA-COSMOS ~\&~COSMOS~collaborations
}

\begin{abstract}
We present results of our 1.4~GHz image stacking analysis of mass-selected galaxies in the COSMOS field. From the
resulting median radio continuum flux density we have determined the evolution of the average star formation rate (SFR) of 
galaxies as a function of stellar mass, unbiased from effects of dust but also source confusion due to the 1.5" angular 
resolution achieved by the VLA. We find a power-law relation between specific SFR (SSFR) and stellar mass for star forming 
galaxies out to $z=3$. While higher mass systems exhibit lower SSFRs at any epoch, no differential, more rapid evolution of 
high mass galaxies is evident at least out to $z \sim 1.5$ where our conclusions are most robust. Utilizing measured mass 
functions of star forming systems, the characteristic stellar mass for galaxies contributing most to the comoving SFR density appears not to evolve. These findings hence challenge 'downsizing' scenarios in which star formation has gradually 
shifted towards lower mass systems with cosmic time. Our analysis constitutes the to-date best determination of the cosmic 
star formation history (CSFH) since $z = 3$ and yields indirect evidence for a rapid decline of the global mass density of 
molecular gas with time. 
\end{abstract}

\section{Introduction}
A complete census of the global cosmic star formation activity at a given cosmic epoch is usually obtained by measuring/
extrapolating the total star formation rate (SFR) per unit comoving volume. The time evolution of this SFR density (SFRD) is 
also known as the cosmic star formation history (CSFH). Over the last years several measurements revealed a steep decline 
of the SFRD since a redshift of $z \sim 1$ and likely a shallower evolutionary behavior at higher redshifts \citep[see][for a 
compilation]{HOPK06}.
Despite the to-date comparatively little observational evidence into the SFRD evolution at high $z$ -- and in particular into 
the fraction of star formation obscured by dust -- it is a widely accepted scenario that the CSFH shows a maximum around $z 
= 2-3$.      

A number of recent studies revealed a surprisingly tight correlation between the total SFR of normal star forming galaxies 
and their stellar mass content \citep[e.g.][for measurements out to $z \sim 1$]{NOES07A, ELBA07}. The localized process of 
star formation is consequently linked to a global galaxy property, suggesting that star formation histories of individual 
galaxies are likely not dominated by stochastic processes such as galaxy mergers \citep[e.g.][]{NOES07A}. The exact 
evolutionary properties of this relation are subject to ongoing debates and insights based on a variety of SFR tracers are 
needed. 

Clearly, multi-wavelength look-back surveys are key to reveal how galaxies build up their stellar mass and how their SFRs 
evolve over a wide range of redshift. The panchromatic datasets from the 2~deg$^2$ cosmic evolution survey (COSMOS; 
\citealp[see][for an overview]{SCOV07B}), in particular, have provided very accurate photometric redshifts (photo-$z$Õs) and 
stellar masses for a large mass-selected sample of galaxies \citep{ILBE09B}.
In this work we use this Spitzer/IRAC ($m_{\rm{AB}}(3.6~\mu \rm{m}) \le 23.9$) selected sample of $\sim$114,000 galaxies in 
combination with high angular resolution Very Large Array (VLA) radio continuum data to measure the evolution of the 
average SFR as a function of stellar mass out to $z \sim 3$. SFRs derived from radio continuum emission require only a simple K-correction and no correction for dust extinction. All methods, results and Figures presented here are discussed in detail in \citet{KARI11}.

\section{The evolution of the specific star formation rate as a function of stellar mass}
Stellar masses have been estimated for all sources via fits of stellar population synthesis models \citep[][BC03]{BRUZ03} to 
the observed spectral energy distributions (SEDs), assuming a \citet{CHAB03} initial mass function, exponentially declining 
star formation histories and a \citet{CALZ00} dust extinction law \citep[details on the estimation of stellar masses and photo-
$z$'s are found in][]{ILBE09B}.
From the restframe intrinsic -- i.e. dust unextincted -- $(\rm{NUV}-r^+)_{\rm{temp}} < 3.5$ colors, obtained from the BC03 SED 
fits, we have separated star forming sources from quiescent ones within our sample. 
 
We have obtained average 1.4~GHz radio flux densities separately for both all and only star forming galaxies from median stacks 
of cutout images of the VLA-COSMOS map (resolution $1.5'' \times 1.4''$; \citealp{SCHI07, SCHI10}) centered at the 
positions of the 3.6~$\mu$m selected sources, after binning each galaxy sample in redshift and mass. We also excluded 
radio-AGN candidates from the sample -- applying various criteria discussed in \citet{KARI11} -- prior to stacking. The 
resulting total flux densities have been converted into SFRs using the calibration of the radio-infrared correlation in the local 
universe \citep{BELL03A}.

\begin{figure*} 
\includegraphics[trim = 0mm 10mm 80mm 1mm, clip,angle=90,width=\textwidth]{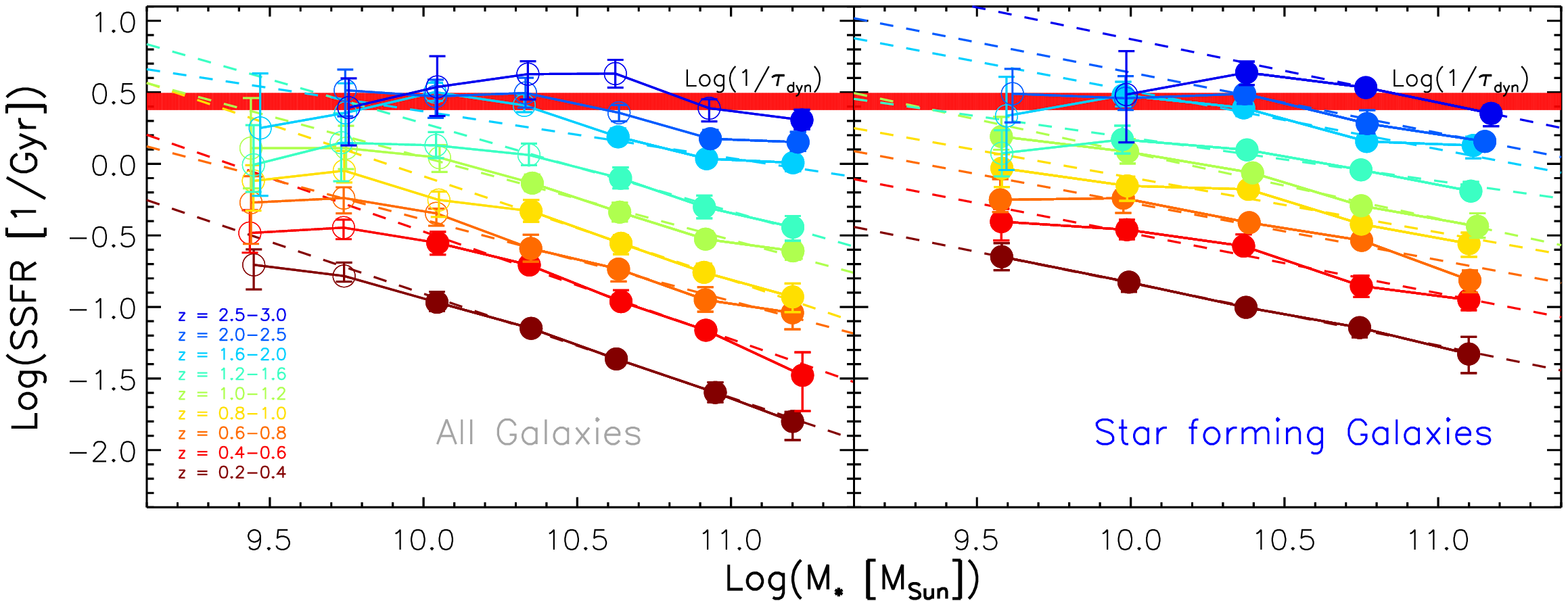} 
\includegraphics[trim = 0mm 10mm 80mm 1mm, clip,angle=90,width=\textwidth]{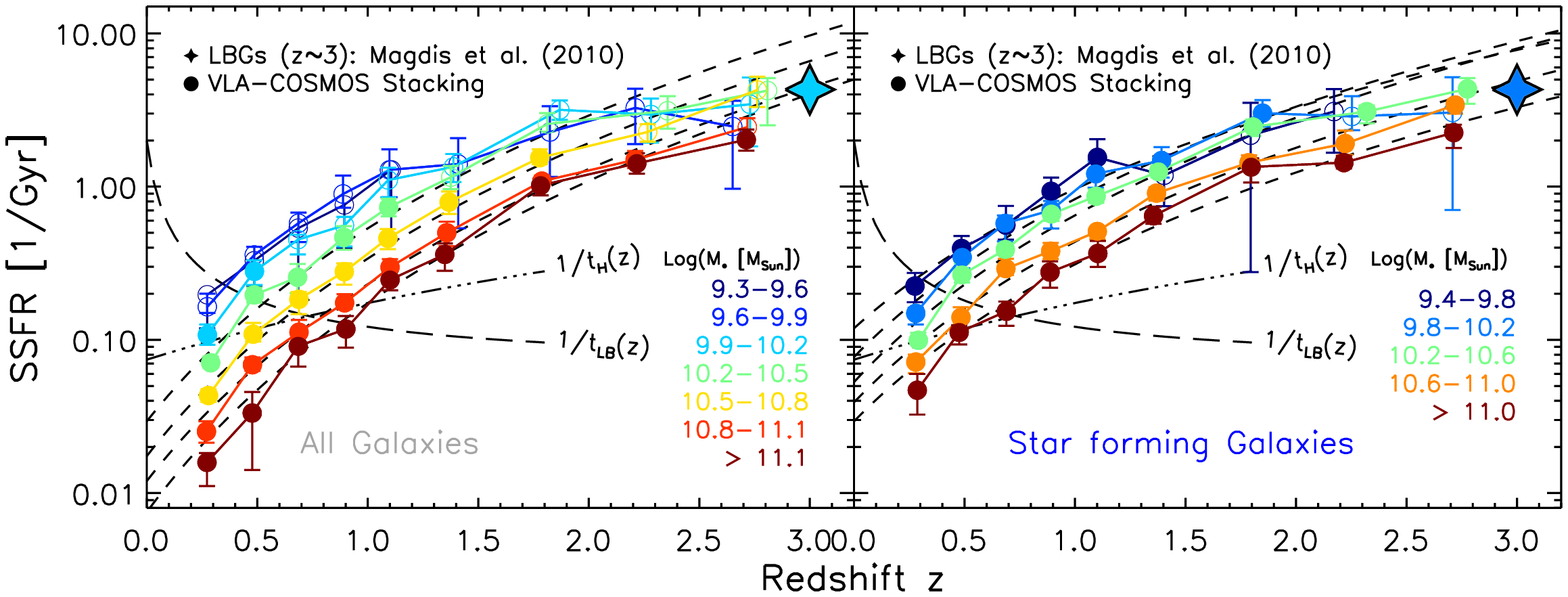} 
\caption{\noindent Radio stacking based measurement of the SSFR as a function of stellar mass at $0.2<z<3.0$ 
(\emph{upper panels}) and redshift evolution of the SSFR in logarithmic stellar mass bins (\emph{lower panels}) for our entire 
galaxy sample (left) and star forming systems only (right). Full symbols depict samples that are regarded as representative for the underlying galaxy population, open symbols represent upper SSFR limits. For representative data points both the 
mass dependence of the average SSFR at a given epoch as well as its redshift evolution at fixed mass follow power-laws 
(dashed lines). Quiescent galaxies (Included in the left panels) are preferentially massive and less frequent at higher $z$ as 
seen in (i) deviations from the high-mass end power-law which itself (ii) becomes gradually slightly shallower with redshift.
For star forming systems only (right panels) the power-law extends over the entire mass range and all mass-bins evolve 
uniformly. A separable power-law $\rm{SSFR}(M_*,z) \propto f(M_*) \times g(z) = M_*^{-0.04} \times (1+z)^{3.5}$ hence is a 
good description for the average SSFR of star forming galaxies at least out to $z \sim 1.5$. The horizontal band 
(\emph{upper panels}) sketches the inverse dynamical timescale, potentially representing an upper bound to the average 
SSFR (see text) and therefore leading to deviations also for high-$z$ star forming sources. The black long-dashed line gives 
the mass-doubling limit above which galaxies are able to double their mass until $z=0$ assuming a constant SFR. The black 
dashed-dotted line depicts the inverse age of the universe at any given redshift and hence makes measured SSFRs 
comparable to the past average star formation activity. The SED-derived measurement of \citet{MAGD10} for LBGs at $z \sim 
3$ with $\log(M_*) \sim 10$ is shown as a filled star.}
\label{fig:ssfrvsmf} 
\end{figure*}

The average stellar mass normalized SFR (specific SFR/SSFR) is best described by separable power-laws in mass and 
redshift for both all (massive) galaxies (Fig. \ref{fig:ssfrvsmf}, left panels) and star forming ones (Fig. \ref{fig:ssfrvsmf}, right 
panels). At $z>1.5$ the average SSFR of low mass star forming galaxies tends to deviate from the mass-uniform 
evolutionary trend seen at lower redshifts. There is thus tentative evidence for an upper limit to the average SSFR at just 
about the inverse of a dynamical timescale of typical local but also higher $z$ disk galaxies \citep[e.g.][]{DADD10}. While 
their dynamical and free-fall times are approximately equal, normal galaxies are assumed to accrete their gas from their 
surroundings and to process it into stars at a constant efficiency \citep[e.g.][]{DADD09}. The upper SSFR-limit we propose 
might hence represent an effective gas accretion threshold.

\section{The characteristic mass of star formation}
\begin{figure*} 
\includegraphics[angle=90,width=\textwidth]{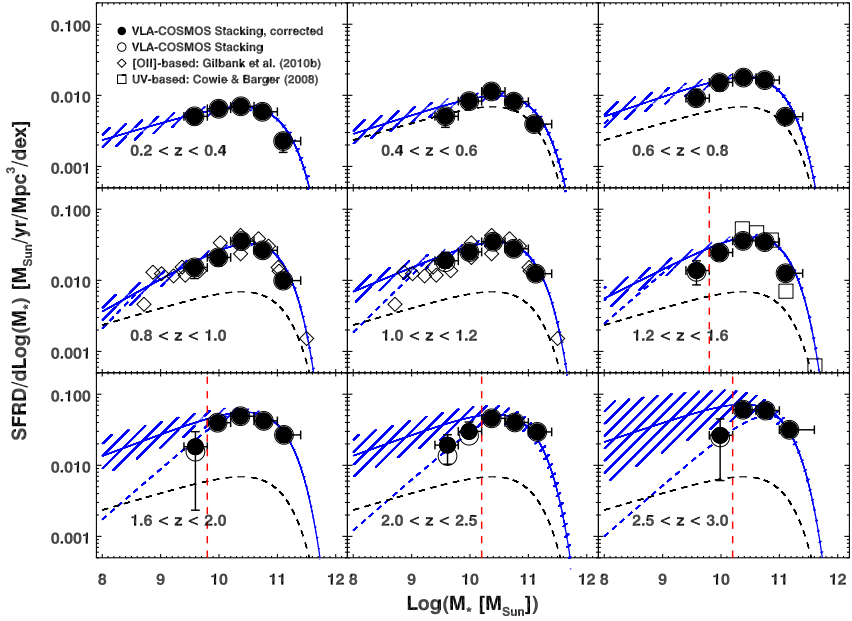} 
\caption{\noindent  The distribution of the SFRD with respect to stellar mass at different epochs out to $z\sim 3$. The data 
points in each panel are the product of the observed number densities of star forming galaxies with the average (stacking-
based) radio SFRs. Leftwards of the dashed vertical lines -- in the mass range where our data are not fully representative for 
the undelying galaxy population -- number densities have been corrected using the corresponding mass functions 
\citep{ILBE09B}. Uncorrected data are shown as open circles. These corrections are generally small and needed only at $z > 
1.5$. In each panel we overplot the Schechter function that results from multiplying the best-fit radio derived SFR-mass 
relation at a given epoch with the corresponding mass function for SF galaxies. Literature results based on the [OII]$
\lambda3727$ line \citep{GILB10A} as well as UV emission \citep{COWI08} agree well with the trends in our data. Dashed 
blue lines show the distribution obtained if an upper limit to the average SSFR at lower masses is assumed.}
\label{fig:sfrdfunc} 
\end{figure*}

While star forming sources exhibit this remarkably shape-invariant SSFR-mass relation throughout cosmic time in our study, it 
is worth noting that their stellar mass functions are found to show a similar constancy of their Schechter profile \citep[e.g.][]
{ILBE09B}. As both phenomena involve a power-law in stellar mass, the product of the SSFR-mass relation and the mass 
function at a given cosmic epoch also yields a Schechter function for the mass distribution of the comoving SFR density 
(SFRD). 
In Fig. \ref{fig:sfrdfunc} we plot the resulting Schechter SFRD functions along with our direct, radio stacking based, 
measurements of the SFRD for each bin in mass and redshift and with literature results from alternative SFR diagnostics at 
different wavelengths. While the potential upper SSFR-limit discussed above might alter the low-mass end shapes 
significantly only at the highest redshifts, the galaxy mass at which the function peaks at {\emph{any}} epoch is $M_*  \sim 
10^{10.6}~ M_{\odot}$, regardless of the assumption of an SSFR-limit. The existence of this characteristic mass of star 
forming galaxies challenges a 'downsizing' paradigm in which the dominant contributors to the CSFH have been massive sources in the cosmic past and low-mass galaxies at present times. It does, however, not rule out that low-mass sources gain more relative importance over time.   

\section{The dust unbiased cosmic star formation history}
\begin{figure*} 
\centering
\includegraphics[width=0.7\textwidth]{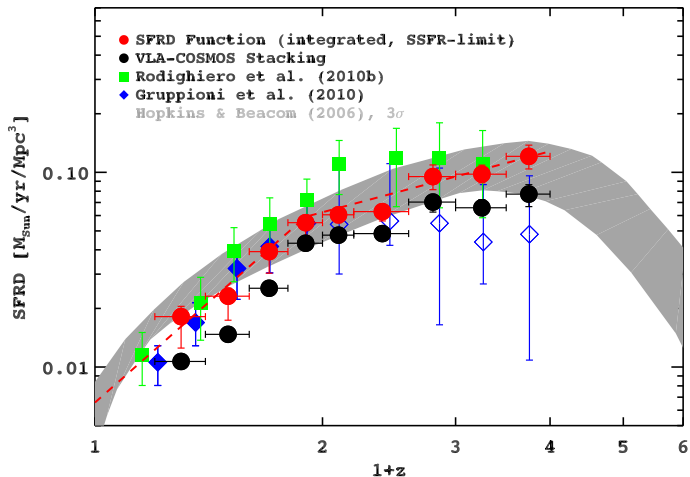} 
\caption{\noindent The dust-unbiased CSFH out to $z=3$. Black circles represent the sum of data points the corresponding 
redshift-bin panel of Fig. \ref{fig:sfrdfunc} and hence a direct -- stacking-based -- measurement of the SFRD down to the 
limiting mass at each epoch. Red filled circles correspond to the 'total' SFRD at each epoch, obtained by integrating the 
Schechter-function fit (Fig. \ref{fig:sfrdfunc}) down to $M_* = 10^5~M_{\odot}$ and assuming an upper limit to the average 
SSFR. The redshift evolution can be described by a broken power-law (dashed lines) that results from the joint 
(non-)evolution of the star forming stellar mass density and the evolution of the SFR-mass relation. 
Recent mid- to far-IR measurements of SFRDs between $0< z <2.5$ (Spitzer/MIPS: \citealp{RODI09}; Herschel/PACS: 
\citealp{GRUP10}) are shown along with our data and the $3\sigma$ envelope from the \citet{HOPK06} compilation. Open 
symbols denote lower limits. Note the remarkable agreement of the far-IR- and radio-based data at all $z$.} 
\label{fig:sfrdall} 
\end{figure*}

The sum of our direct radio stacking based measurements of the SFRD for individual mass bins yields lower limits to the total 
SFRD at any redshift. At a given redshift the Schechter SFRD function enables us to obtain the remainder of the total SFRD, 
integrated below our observational mass limits. For this integration we assume the upper bound to the average SSFR 
motivated above. The resulting radio-based -- i.e. dust-extinction unaffected -- CSFH out to $z \sim 3$ is shown in Fig. 
\ref{fig:sfrdall} along with independent literature measurements.
Since $z \sim 1$ the stellar mass density of star forming galaxies is approximately constant \citep[e.g.][]{ILBE09B} and also 
the relative fractions of objects at different masses within this population are conserved due to the shape-invariance of their 
mass function (see above). In this redshift range the CSFH is therefore entirely controlled by the mass-uniform evolution of 
the average SFR. Given a constant star formation efficiency, a strong and global decline in the mass density of molecular gas 
thus entirely explains the observed decrease of the integrated SFRD with cosmic time. Towards earlier epochs it is predominantly the -- again mass-uniform -- decrease in star forming stellar mass density that leads to a shallower evolutionary behavior. The CSFH since $z \sim 3$ is hence well described by a broken power-law (Fig. \ref{fig:sfrdall}).

\section{Summary and conclusions}
Using a median image stacking technique of 1.4~GHz radio continuum emission and an unprecedentedly rich sample of 
galaxies selected at 3.6~$\mu$m with panchromatic (FUV to mid-IR) ancillary data in the COSMOS field we have measured 
stellar mass-dependent average (specific) star formation rates ((S)SFRs) in the redshift range $0.2 < z < 3$. We have applied 
various criteria to minimize contaminating radio flux from potential active galactic nuclei and have used the template-based, 
intrinsic rest-frame $(\rm{NUV}-r^+)_{\rm{temp}}$ color from SED-fits in the NUV-mid-IR in order to separate star forming 
sources from quiescent systems. Our findings -- presented in detail in \citet{KARI11} -- are:
\begin{itemize}
\item The evolution of the average (S)SFR of stellar mass selected galaxies is approximately mass-independent at least out 
to $z \sim 1.5$ where the depth of our data allows the most robust conclusions. A power-law relation between (S)SFR and 
mass represents well the data at the high mass end. 
\item The (S)SFR of star forming sources at $z<3$ also evolves basically independent of mass while a power-law 
dependence of (S)SFR and mass generally extends over all masses probed. Higher mass galaxies 
hence have lower SSFRs, regardless if quiescent galaxies are included in the analysis or not.
\item There is tentative evidence for an upper bound to the average SSFR preventing a further growth of the SSFR with 
$z$, starting at the low-mass end at $z \sim 1.5-2$.
\item The stellar mass distribution function of the comoving SFR density at any epoch is well described by a 
Schechter function that reveals a constant characteristic mass of star formation since $z \sim 3$. Its integral yields the total 
SFRD at any $z$.
\end{itemize}

\acknowledgements We thank A. Mart\'\i nez-Sansigre, M. Sargent, H.-W. Rix, A. van der Wel, O. Ilbert, V. Smol{\v{c}}i{\'{c}}, C. Carilli. M. Pannella, A. Koekemoer, E. Bell and M. Salvato for their numerous contributions to this work.

\bibliography{karim_guilin}

\end{document}